\documentclass[12pt]{iopart}
\usepackage{iopams}  
\usepackage{epsf}
\begin{document}

\title[Polymer adsorption transition]
{Adsorption transition of a self-avoiding polymer 
chain onto a rigid rod}

\author{Andreas Hanke} 

\address{Department of Physics, University of
Texas at Brownsville, Brownsville, TX 78520, USA}

\ead{hanke@phys.utb.edu}

\begin{abstract}
The subject of this work is the adsorption transition 
of a long flexible self-avoiding polymer 
chain onto a rigid thin rod. 
The rod is represented by a cylinder of radius $R$ 
with a short-ranged attractive surface potential
for the chain monomers. General scaling results are 
obtained by using renormalization group arguments in 
conjunction with available results for quantum field 
theories with curved boundaries
[McAvity and Osborn 1993 {\em Nucl. Phys. B} 
{\bf 394} 728].
Relevant critical exponents are identified and 
estimated using geometric arguments.
\end{abstract}


\vspace*{15mm}

\noindent
It is a pleasure to dedicate this work to
L.~Sch\"afer on the occasion of his 60th birthday.

\maketitle

\section{Introduction}

Polymers and polymer solutions belong to the most 
intensively studied objects in condensed matter 
physics \cite{dG79,CJ90,LS99}.
The adsorption of polymers on surfaces and interfaces 
is of special importance \cite{Fleer}. 
Adsorption of free polymers in solution on 
the container wall or other boundaries occurs in
the presence of attractive interactions between 
the surface and the chain monomers. Examples of such 
interactions include Coulomb and van der Waals forces, 
and more specific molecular interactions. Coulomb 
forces are screened by counter-ions in the solution
and can be tuned to some extent by adding salt to 
the solvent. Thus on changing the properties of the 
solvent an individual polymer chain can undergo a 
reversible transition from a freely floating, desorbed 
state to an adsorbed state in which the chain monomers 
are close to the surface on average. The adsorption of 
polymers on flat surfaces has been studied theoretically 
and experimentally, and is by now well understood
\cite{dG79,Fleer,EKB82,PNS}. 
Due to the importance to colloidal dispersions the 
interaction of polymers with spherical and rodlike 
particles has been studied as well 
\cite{Napper,Buit,PVL95,EHD,Now}. The adsorption of 
flexible polymers on rodlike particles is relevant, for 
example, in gels \cite{PVL95}, and for the binding of 
flexible polymers to relatively stiff biomolecules such 
as DNA \cite{Alberts}. Another class of polymer 
adsorption transitions involve two flexible 
self-avoiding but mutually attracting polymers which
can form a bound, double-stranded, so-called ``zipped''
state. A prominent example of this kind of
transition is the denaturation transition of 
double-stranded DNA \cite{PS,WB} which recently
attracted considerable attention regarding its
statistical-mechanical properties and 
thermodynamic order 
\cite{CCG00,KMP2000,GMO01,COS02,RG04,Schaefer}
(for a recent review on the biophysics related to 
DNA topology, see \cite{MH05}).
The DNA denaturation transition is usually modelled 
in such a way that monomer $i$ of one strand
can only interact with monomer $i$ of the other
strand, reflecting the key-lock principle of natural, 
inhomogeneous DNA with its specific sequence 
of base pairs.
Two-chain systems in which any monomer of one
chain can interact with any monomer of the other
chain include diblock copolymers, which consist of
linear chains of $N$ monomers of type $A$ followed 
by $N'$ monomers of type $B$, with different $AA$, 
$BB$, and $AB$ interactions; systems of this kind
have been intensively studied as well
\cite{JBL84,SK85,SK90,SLK91,FH97,V98}.
Recently it was found
that self-avoiding mutually attracting diblock 
copolymers can adopt a zipped state in which the 
two components form a bound, double-stranded 
structure which however remains 
swollen and does not assume compact configurations.
The zipped state is located between a swollen, 
unbound high-temperature state and a compact 
low-temperature state, separated by second-order
and first-order phase transitions, respectively
\cite{OSS2000,BCOS01,KGB04}.

\begin{figure}
\unitlength=1cm 
\begin{picture}(10,8) 
\put(-0.5,-30.5){\includegraphics{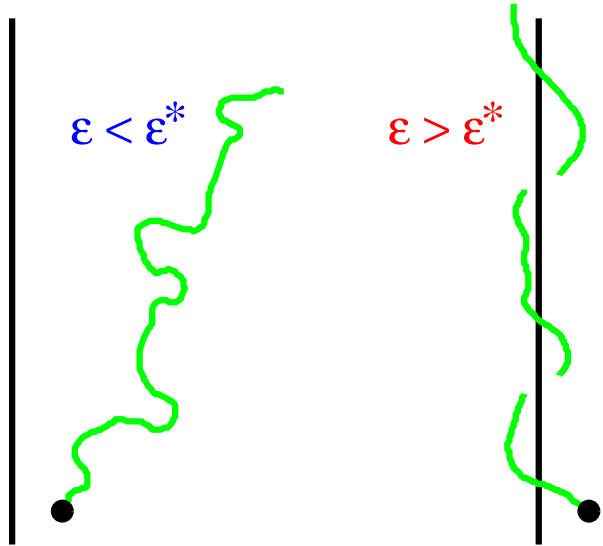}} 
\end{picture}
\caption{A long flexible self-avoiding
polymer chain (green) adsorbs onto 
a thin elongated rod (black) if the 
adsorption energy per chain monomer
${\cal E}$ is larger than the 
threshold value ${\cal E}^*$.}
\label{fig_rod}
\end{figure} 

In this work we focus on the adsorption transition of a 
long flexible self-avoiding polymer chain onto a rigid 
thin elongated rod, as shown in figure \ref{fig_rod}. 
We assume that the rod is endowed with a short-ranged 
surface potential, or adsorption energy, ${\cal E}$,
for the chain monomers; 
the more positive ${\cal E}$,
the more attractive the interaction.
Thus, on increasing ${\cal E}$ from a low value,
at some threshold value ${\cal E}^*$ the chain 
is expected to undergo a transition from an unbound, 
free state to a bound state in which the overall 
gain in binding energy compensates the loss of 
configurational entropy. An interesting feature
of this transition is the fact that it 
represents a true phase transition in the 
thermodynamic sense; the adsorbed state forming an 
elongated, aligned structure, see figure \ref{fig_rod}.
In contrast, for a particle of finite size,
a self-avoiding polymer chain cannot undergo a true
adsorption phase transition due to steric constraints.
The focus of the present work is on the former case. 
Since the polymer adsorption transition is 
characterized by large fluctuations we 
expect scaling and universal behaviour.
We thus use the renormalization group to obtain 
general scaling results for the chain partition function.
We also obtain estimates for relevant 
critical exponents by geometric arguments.
Before we introduce and study our model in 
section \ref{sec_rod} we recall 
some general ideas and concepts for later reference. 
In section \ref{subsec_general} we discuss 
typical scaling arguments related to the polymer 
adsorption transition. Since our work strongly relies 
on field-theoretical methods, in section 
\ref{subsec_map} and \ref{sec_app} we sketch 
the mapping of the polymer system on the 
Ginzburg-Landau model.


\subsection{General scaling behaviour}
\label{subsec_general}

Consider the adsorption of a long flexible 
polymer chain onto an object ${\cal S}$. 
For the time being,
this object can be a surface, a thin rod, 
another flexible polymer chain, or any other 
extended manifold that allows for a thermodynamic 
adsorption phase transition. The quantity 
of interest is the 
partition function $Z$ of the chain 
with one end fixed close to the object ${\cal S}$ and 
the other end moving freely. 
Upon adjusting the system's thermodynamic variables 
close to the adsorption transition point, 
only the number of chain monomers $N \gg 1$ and 
the adsorption energy ${\cal E} \approx {\cal E}^*$ 
remain as relevant parameters, where ${\cal E}^*$
is the adsorption energy at the transition point.
The partition 
function $Z$ is expected to obey the scaling
\begin{equation} \label{scale}
Z(N, {\cal E}) \sim p^N N^{\gamma' - 1} 
f[({\cal E}- {\cal E}^*) N^{\Phi}]
\end{equation}
where $p$ is the lattice-dependent connectivity 
constant and $\gamma'$, $\Phi$ are critical 
exponents. The scaling function $f(x)$ is regular 
at $x = 0$ since $Z(N, {\cal E})$ has no singularity 
for finite $N$ and ${\cal E} \approx {\cal E}^*$. 
The exponent $\gamma'$ thus characterizes the scaling of 
$Z$ right at the transition point:
$Z(N,{\cal E}^*) \sim p^N N^{\gamma' - 1}$.
Note that $\gamma'$ is not necessarily equal to the 
critical exponent $\gamma$
for an unbounded, free chain, 
for which $Z_{free}(N) \sim p^N N^{\gamma - 1}$
(compare equation (\ref{scalbehz}) in section
\ref{subsec_ren}, with $\gamma_1$
introduced in equation (\ref{scalbeh}) and $L \sim N$).
The exponent $\Phi$ is referred to as the crossover-exponent.
Since $- {\cal E}$ acts as a chemical potential for 
monomers close to ${\cal S}$, the number 
$N_S$ of adsorbed monomers scales as
\begin{equation} \label{ns}
N_S \sim \frac{d}{d {\cal E}} \ln Z(N, {\cal E}) \, \, .
\end{equation}
Equation (\ref{scale}) implies three 
distinct scaling regimes for $N_S$.

(i) ${\cal E} = {\cal E}^*$. Equations (\ref{scale}) 
and (\ref{ns}) yield
\begin{equation} \label{nscale}
N_S \sim N^{\Phi}  \, \, , \qquad
{\cal E} = {\cal E}^* \, \, , \, \, N \to \infty \, \, .
\end{equation}
For $0< \Phi < 1$ this implies that $N_S$ grows with $N$
but the {\em fraction} of adsorbed monomers, $N_S / N$, vanishes 
for $N \to \infty$. For $\Phi = 1$, the behaviour $N_S \sim N$
at ${\cal E} = {\cal E}^*$ indicates that the adsorption 
transition is of first order. 

(ii) ${\cal E} < {\cal E}^*$. Equation (\ref{scale})
implies that the scaling behaviour of $Z$ for 
$N \to \infty$ is governed by the behaviour of $f(x)$ 
for $x \to - \infty$, regardless the precise value 
of ${\cal E}$. In this case 
${\cal S}$ is repulsive for the chain monomers 
and $N_S$ stays finite even for $N \to \infty$.

\begin{figure}
\unitlength=1cm 
\begin{picture}(10,5.2) 
\put(-2.7,-36.2){\includegraphics{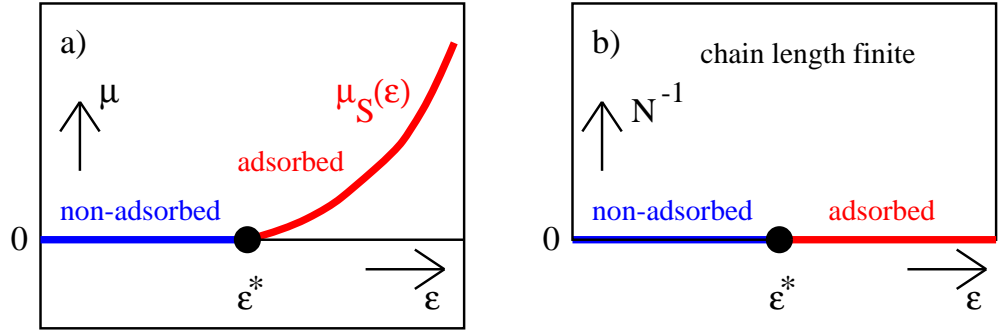}} 
\end{picture}
\caption{Typical phase diagrams for polymer adsorption
in the a) grand canonical ensemble $(\mu, {\cal E})$
and b) canonical ensemble $(N, {\cal E})$. 
The manifold
${\cal S}$ on which the polymer adsorbs can be any 
object which allows for an adsorbed thermodynamic state.
For $N \to \infty$, the fraction of 
adsorbed monomers $N_S / N$ vanishes for 
${\cal E} < {\cal E}^*$ (blue lines)
whereas it is finite for ${\cal E} > {\cal E}^*$ 
(red lines).
The shape of the red line in a) is given by the function
$\mu_S({\cal E})$ in equation (\ref{s}).
Compare figure \ref{fig_rod}.}
\label{fig_pd}
\end{figure}

(iii) ${\cal E} > {\cal E}^*$. The chain adopts 
an adsorbed state and stays close to ${\cal S}$ 
on average. Thus, $N_S \sim N$, which implies
a {\em finite} fraction of adsorbed monomers 
for $N \to \infty$:
\begin{equation} \label{fraction}
F({\cal E}) \equiv \lim\limits_{N \to \infty} \frac{N_S(N,{\cal E})}{N}
> 0 \, \, \, , \qquad {\cal E} > {\cal E}^* \, \, \, .
\end{equation}
To analyze the behaviour of $F({\cal E})$ it is 
useful to consider the grand canonical ensemble.
The partition function in the grand canonical ensemble,
${\cal X}(\mu, {\cal E})$, is related to $Z(N, {\cal E})$ 
by a Laplace transform: 
\begin{equation} \label{grand}
{\cal X}(\mu, {\cal E}) = \int_0^{\infty} d N 
e^{- \mu N} Z(N, {\cal E}) \, \, ,
\end{equation}
where 
$\mu$ is the chemical potential conjugate to $N$.
Equation (\ref{grand}) is valid for 
$\mu > \mu_c$ with $\mu_c = \ln p$. 
One is allowed to set $p = 1$ for simplicity,
so that $\mu_c = 0$. Equation (\ref{scale}) 
then implies the scaling behaviour
\begin{equation} \label{scalex}
{\cal X}(\mu, {\cal E}) \sim \mu^{-\gamma'}
g[({\cal E}- {\cal E}^*) \mu^{- \Phi})] \, \, \, , 
\qquad \mu > 0 \, \, .
\end{equation}
By reasoning similar to that below equation
(\ref{scale}) one finds that 
the scaling function $g(y)$ is 
regular at $y = 0$ and $\gamma'$
characterizes the scaling of ${\cal X}$ right 
at the transition point:
${\cal X}(\mu, {\cal E}^*) \sim \mu^{-\gamma'}$.
On the other hand, we 
know that for ${\cal E} > {\cal E}^*$ the chain 
takes an adsorbed state. 
For the grand canonical ensemble this
implies, for given ${\cal E} > {\cal E}^*$, 
that $N_S \to \infty$ for
$\mu \searrow \mu_S({\cal E})$ with some
$\mu_S({\cal E}) > 0$. 
In this limit we thus expect the scaling behaviour 
${\cal X}(\mu, {\cal E}) \sim (\mu - \mu_S)^{-\gamma_S}$
where $\gamma_S$ is characteristic for the {\em adsorbed} 
state and {\em different} from $\gamma'$.
For example, if ${\cal S}$
is another flexible polymer chain,
the adsorbed state forms a double-stranded structure
which, as a whole, behaves like an unbounded, free,
self-avoiding chain, 
which implies $\gamma_S = \gamma$ in this case \cite{CCG00}.
Using equation (\ref{scalex})
it follows that the scaling function $g(y)$ must have a 
singularity at some $y_S > 0$ of the form
\begin{equation} \label{sing}
g(y \nearrow y_S) \sim (y_S - y)^{-\gamma_S} \, \, .
\end{equation}
The relation 
$({\cal E}- {\cal E}^*) \mu_S^{- \Phi} = y_S$
determines the shape of the line $\mu_S({\cal E})$ as
\begin{equation} \label{s}
\mu_S({\cal E}) \sim ({\cal E} - {\cal E}^*) ^{1/\Phi} 
\, \, \, , \quad {\cal E} > {\cal E}^* \, \, \, .
\end{equation}
Figure \ref{fig_pd} shows typical phase diagrams for polymer 
adsorption in the grand canonical ensemble (fixed $\mu$) 
and canonical ensemble (fixed $N$).

According to the above, 
for ${\cal E} > {\cal E}^*$, ${\cal X}(\mu, {\cal E})$ in 
equation (\ref{grand}) has a singularity 
for $\mu \searrow \mu_S({\cal E}) > 0$, and $N$ $(\ge N_S)$ 
diverges in this limit. This, in turn, implies
\begin{equation} \label{divz}
Z(N, {\cal E}) \sim \exp[\mu_S({\cal E}) N] \, \, , \qquad
{\cal E} > {\cal E}^* \, \, , \, \, N \to \infty \, \, .
\end{equation}
Using equation (\ref{ns}) one finds
\begin{equation} \label{resns}
N_S \sim N \frac{d}{d {\cal E}} \, \mu_S({\cal E}) \, \, \, ,
\end{equation}
and thus, using equations (\ref{fraction})
and (\ref{s}),
\begin{equation} \label{resfracintro}
F({\cal E}) = \lim\limits_{N \to \infty} \frac{N_S}{N}
\sim \frac{d}{d {\cal E}} \, \mu_S({\cal E})
\sim ({\cal E} - {\cal E}^*)^{\kappa} \, \, \, , 
\quad {\cal E} > {\cal E}^* \, \, \, ,
\end{equation}
where the exponent $\kappa$ is related to the 
crossover-exponent $\Phi$ in equation (\ref{nscale})
by
\begin{equation} \label{relation}
\kappa = \frac{1 - \Phi}{\Phi} \, \, .
\end{equation}
In particular, for $\Phi = 1$ the fraction 
$\lim\limits_{N \to \infty} N_S / N$ 
jumps from zero for 
${\cal E} < {\cal E}^*$ to a finite value 
for ${\cal E} > {\cal E}^*$, which is then independent of 
${\cal E}$; this corresponds to a first-order
transition (compare case (i) above). 

The scaling behaviours (\ref{nscale}) and 
(\ref{resfracintro}), (\ref{relation}) have been proven
rigorously for the polymer adsorption transition
on a {\em flat} surface \cite{EKB82,PNS}.
This system is closely related to the semi-infinite
Ginzburg-Landau model, see \cite{Binder,Diehl,Diehl2} 
for reviews;
the mapping of the polymer system on the 
Ginzburg-Landau model is discussed below. 
The scaling behaviours (\ref{nscale}) and 
(\ref{resfracintro}), (\ref{relation}) also hold
reasonably well in a recent numerical study of
the DNA denaturation transition \cite{CCG00}.


\subsection{Mapping of the polymer system on the 
Ginzburg-Landau model} \label{subsec_map}

According to Edwards' continuous chain model we represent the 
configuration of a linear chain of length $L$ by a curve 
${\bf R}(s)$, parameterized by its arc length $s$,
in $D$-dimensional space. The chain length $L$ is proportional
to the number of chain monomers $N$. In the presence of an
external potential $V({\bf r})$ the partition function of 
the chain is given by
\begin{eqnarray} \label{pf}
Z^{(2)}({\bf r}, {\bf r}'; L) & = &  
\int_{\bf r}^{{\bf r}'} {\cal D} {\bf R}
\, \exp \left\{ - \frac{1}{4}
\int_0^L ds \left(\frac{d {\bf R}}{ds}\right)^2 \right\} \\[2mm]
& & \times \, \exp \left\{-
\int d^Dr \left[V({\bf r}) \rho({\bf r})  
+ \frac{u}{6} \, \rho^2({\bf r}) \right] \right\}
\nonumber \end{eqnarray}
with the monomer density
\begin{equation}
\rho({\bf r}) = \int_0^L ds \, \delta^D( {\bf r} - {\bf R}(s)) \, \, .
\end{equation}
The symbol $\int_{\bf r}^{{\bf r}'} {\cal D} {\bf R}$
denotes functional integration over all chain configurations
with the chain ends fixed at ${\bf r}$ and ${\bf r}'$. 
The superscript ``$(2)$'' on $Z$
indicates that the chain is fixed with
both ends. 
The coupling constant $u$ of the $\rho^2({\bf r})$ 
interaction characterizes the strength of the contact 
interaction between chain monomers: $u = 0$ 
describes a Gaussian random walk whereas $u > 0$
describes a self-avoiding chain.
The case $u < 0$ is related to the polymer 
collapse transition to a compact state in a poor 
solvent \cite{dG75,Dup82}; in this work we do not 
consider this collapse transition, 
and therefore exclude the case $u < 0$. 

As first noticed by de Gennes \cite{dG72},
the polymer system can be mapped on the 
Ginzburg-Landau model of an $n$-component order 
parameter field $\vec{\Phi} = (\Phi_1, \ldots, \Phi_n)$ 
in the limit $n \to 0$. 
It is worthwhile to note that this mapping not 
only works in perturbation theory but already on the 
level of the Hamiltonian in the Ginzburg-Landau model.
The derivation, using a Gaussian transformation 
to linearize the $\rho^2({\bf r})$ interaction 
in equation (\ref{pf})
\cite{C75}, is left to \ref{sec_app}.
The result is
\begin{equation} \label{res}
Z^{(2)}({\bf r}, {\bf r}'; L) = 
{\cal L}_{t \to L} \, \lim\limits_{n \to 0} \,
\langle \Phi_1({\bf r}) \, \Phi_1({\bf r}') \rangle
\end{equation}
where 
\begin{equation} \label{tpf}
\langle \Phi_1({\bf r}) \, \Phi_1({\bf r}') \rangle
= \int {\cal D} \vec{\Phi} \,
\Phi_1({\bf r}) \, \Phi_1({\bf r}') \, e^{- {\cal H}\{\vec{\Phi}\}}
\end{equation}
is the two-point correlation function in the
Ginzburg-Landau model with the standard Hamiltonian
\cite{Amit,ZJ02}
\begin{equation} \label{action}
{\cal H}\{\vec{\Phi}\} = \int d^Dr \left[ 
\frac{1}{2} \, (\nabla \vec{\Phi})^2 + 
\frac{t}{2} \, \vec{\Phi}^2 + 
\frac{1}{2} \, V({\bf r}) \, \vec{\Phi}^2 +
\frac{u}{24} \, (\vec{\Phi}^2)^2 \right] \, \, .
\end{equation}
The operation
\begin{equation} \label{laplace}
{\cal L}_{t \to L} = \frac{1}{2 \pi i} \int_{\cal C} dt \,
e^{t L} 
\end{equation}
acting on the correlation function in equation (\ref{res})
is an inverse Laplace transform 
in which the integration path ${\cal C}$ in the complex 
$t$-plane is a parallel to the imaginary 
axis to the right of all singularities.
Equations (\ref{res}) - (\ref{laplace}) describe
the statistics of the polymer chain in terms of
properties of the near-critical ferromagnetic $n$-vector 
model in the limit $n \to 0$. In the context of the 
$n$-vector model, the
parameter $t \sim T - T_c$ describes the deviation of the
temperature $T$ from the critical temperature $T_c$.
The form of the interaction involving the potential
$V({\bf r})$ in equation (\ref{action}) shows that
the $O(n)$-invariant scalar $\vec{\Phi}^2$ is related to 
the monomer density $\rho({\bf r})$ 
of the polymer chain in equation (\ref{pf}).
Thus, translated to the polymer system, the term
$\vec{\Phi}^2 \cdot \vec{\Phi}^2$ is related to the contact 
interaction between chain monomers and $t$ plays the role 
of a chemical potential for chain monomers in the bulk.


\section{Polymer adsorption transition onto a rigid rod} 
\label{sec_rod}

\begin{figure}
\unitlength=1cm 
\begin{picture}(10,9) 
\put(1,-25){\includegraphics{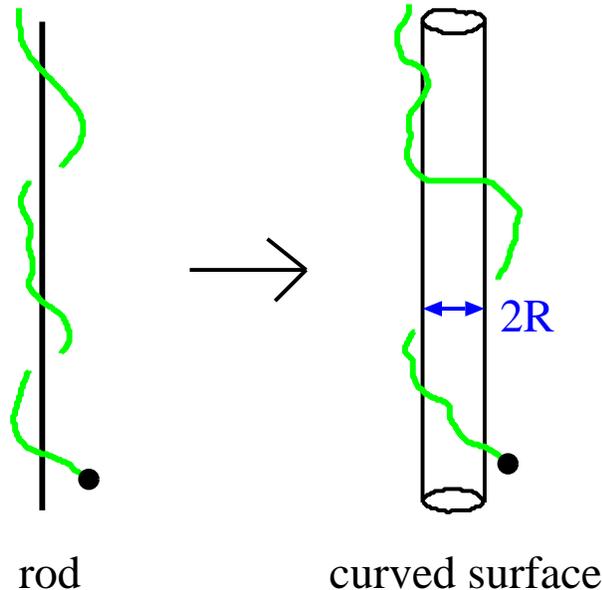}} 
\end{picture}
\caption{The rod is represented by a cylinder with finite 
radius $R$. This allows one to use available 
results for field theories with curved boundaries, but 
introduces the additional variable $R$ into the problem.
Compare figure \ref{fig_rod}.}
\label{fig_curved}
\end{figure}

The objective of this work is the study of the
adsorption of a long 
flexible polymer chain onto a rigid thin rod. However,
for the time being we model the rod by an infinitely 
elongated cylinder with small but finite radius $R$. 
The introduction of a finite cylinder radius $R$ 
is necessary since the limit of a rod with zero radius 
turns out to be too singular for the present 
field-theoretical treatment; see figure 
\ref{fig_curved} and the discussion below equation 
(\ref{fracd2}). On the other hand, the fact that the 
adsorption transition now takes place on a {\em surface}, 
albeit a curved one, allows us to take advantage
of available results for field theories with curved 
boundaries \cite{AO93}; see section \ref{subsec_ren} below.
The chain of total length $L_0 \sim N$, where $N$ is the
number of chain monomers,
is fixed with one end at the point ${\bf r}_S$ on 
the cylinder surface $S$ while the other end is moving
freely. The cylinder surface is endowed with a short-ranged 
surface potential $c_0$ acting on the chain monomers, and it 
is understood that the chain monomers are excluded from the 
interior of the cylinder. The potential V({\bf r}) in 
equation (\ref{pf}) is thus given by
\begin{equation} \label{vs}
V({\bf r}) = c_0 \int_S dS' \, \delta^D({\bf r} - {\bf r}'_S)
\end{equation}
and $V({\bf r}) = \infty$ if ${\bf r}$ is located in the interior
of the cylinder. By virtue of equation (\ref{res}), the chain 
partition function is given by
\begin{equation} \label{intz}
Z(L_0) = \int_V d^D r' \, Z^{(2)}({\bf r}_S, {\bf r}'; L_0)
= {\cal L}_{t_0 \to L_0} \, \lim\limits_{n \to 0} \,
\chi(t_0) \, \, .
\end{equation}
The integration volume 
$V$ is the outer space of the cylinder bounded 
by the cylinder surface $S$. On the right hand side,
$\chi(t_0) \equiv \chi({\bf r}_S; t_0)$, where the
{\em susceptibility} $\chi({\bf r}; t_0)$ in the 
Ginzburg-Landau model is obtained by integrating the 
two-point correlation function, i.e., 
\begin{equation} \label{intc}
\chi({\bf r}; t_0) = 
\int_V d^D r' \, \langle \Phi_1({\bf r}) \, 
\Phi_1({\bf r}') \rangle \, \, ,
\end{equation}
in a Ginzburg-Landau type field theory with Hamiltonian
\cite{Binder,Diehl,Diehl2}
\begin{equation} \label{ham}
{\cal H}\{\Phi\} = \int\limits_V d^Dr \left[ 
\frac{1}{2} \, (\nabla \vec{\Phi})^2 + 
\frac{t_0}{2} \, \vec{\Phi}^2 +
\frac{u_0}{24} \, (\vec{\Phi}^2)^2 \right]
+ \int\limits_S dS \,
\frac{c_0}{2} \, \vec{\Phi}^2
\end{equation}
of an $n$-component order parameter field
$\vec{\Phi} = (\Phi_1, \ldots, \Phi_n)$. 
In the following we will understand $\chi({\bf r}; t_0)$
as the chain partition function in the grand canonical 
ensemble, where $t_0$ is conjugate to $L_0$ (compare
section \ref{subsec_general} with
$t_0 \sim \mu$, $L_0 \sim N$, and 
$c_0 \sim - {\cal E}$).
The subscript ``0'' on $c_0$, $t_0$, $L_0$, $u_0$ 
is used to 
distinguish these parameters from their renormalized 
counterparts that will be introduced below.
If ${\bf r} = {\bf r}_S$ in $\chi({\bf r}; t_0)$ 
we suppress the argument ${\bf r}_S$ since $\chi$ 
does not depend of the choice of ${\bf r}_S$ by 
symmetry; compare equation (\ref{intz}).

\begin{figure}
\unitlength=1cm 
\begin{picture}(10,15) 
\put(-2,-18.5){\includegraphics{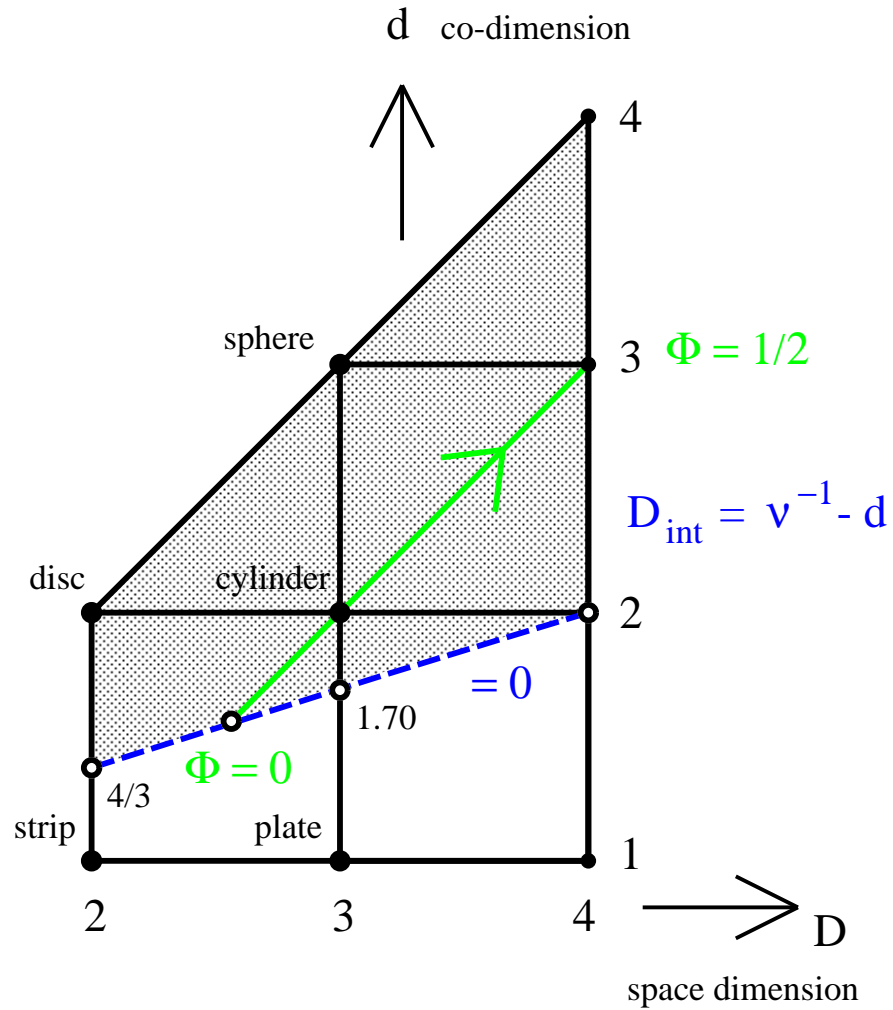}} 
\end{picture}
\caption{Diagram of ``generalized cylinders'' which behave as 
relevant or irrelevant perturbations for long, flexible, 
self-avoiding polymer chains.
The parameter $d \le D$ characterizes the shape of the body and $D$
is the space dimension. The point $(d,D) = (2,2)$  corresponds
to a disc in $D=2$ and the points $(3,3)$ to a sphere in $D=3$.
The point $(2,3)$ corresponds to an infinitely elongated cylinder
in $D = 3$ (in which we are ultimately interested).
The line $D = 4$ indicates the upper critical 
dimension where the polymer chains behave like Gaussian
random walks. The open circles and the dashed line (blue) indicate
points $(d,D)$ for which the dimension $D_{int} = \nu^{-1}(D)-d$
of intersection points of the body with an unbounded, free,
self-avoiding random walk vanishes.
For $D_{int} < 0$ the intersection probability is zero. 
In this sense,
in the shaded region above the dashed line the body
represents an irrelevant perturbation.}
\label{fig_cyl}
\end{figure}

Before we proceed with the renormalization of the model 
defined by equation (\ref{ham}) we review some cases
in which results are available. To this end it is useful 
to consider not only a cylinder in $D = 3$ dimensions
but bodies of more general shape. 
The ``generalized cylinders''
have an infinitely extended axis of dimension $D - d$ and 
a curved surface with constant curvature radius $R$ 
in the subspace of co-dimension $d$.
\footnote{The co-dimension of an object of dimension $D_{obj}$
in a space of dimension $D$ is given by $D - D_{obj}$.}
The axis
can be the axis of an ordinary cylinder in three
dimensions, for which 
$(d,D) = (2,3)$, the midplane of a plate $(d = 1)$, 
or the centre of a sphere $(d = D)$. The explicit 
form of a ``generalized cylinder'' is given by the set
$\left\{ {\bf r} = ({\bf r}_{\perp},{\bf r}_{\parallel}) 
\in {\mathbb R}^{d} \times 
{\mathbb R}^{D-d} ; |{\bf r}_{\perp}| \le R \right\}$.
Figure \ref{fig_cyl} shows some examples
in the $(d,D)$-plane; compare also references 
\cite{EHD,HD99,H2000}.

\subsection{Flat surface}

First, consider the line $d = 1$ with $D$ arbitrary
in figure \ref{fig_cyl}. Then,
of course, equation (\ref{ham}) corresponds to the 
semi-infinite $n$-vector model bounded by a {\em flat} 
surface for which many results are available 
\cite{Binder,Diehl,Diehl2}.
In particular, for the polymer case $n = 0$, the 
considerations of section \ref{subsec_general} can be 
made explicit, proving equations (\ref{nscale}) and
(\ref{resfracintro}), (\ref{relation}) \cite{EKB82,PNS}.


\subsection{Gaussian model} 
\label{subsec_gauss}

Next, consider the line $D = 4$ with $d$ arbitrary
in figure \ref{fig_cyl}.
Since $D = 4$ is the upper critical dimension of the
model in equation (\ref{ham}), this 
case corresponds to the Gaussian field theory and to 
Gaussian random walks, respectively. Both 
cases are described by $u_0 = 0$ in equation (\ref{ham})
and can be solved exactly by standard methods \cite{EHD}.
Thus, the susceptibility defined in equation (\ref{intc}),
corresponding to the chain partition function in the
grand canonical ensemble, is given by \cite{EHD}
\begin{equation} \label{suscept}
\chi({\bf r}; t_0) =
\frac{1}{t_0} \left[1 - \frac{\zeta_0 \, \rho^{- \alpha} 
K_{\alpha}\left(\rho \sqrt{\tau_0}\,\right)}
{\sqrt{\tau_0} \, K_{\alpha+1}\left(\sqrt{\tau_0} \, \right)
+ \zeta_{0\,} K_{\alpha}\left(\sqrt{\tau_0} \, \right) } \right]
\end{equation}
where $\rho = |{\bf r}_{\perp}| / R$ 
(so that $\rho= 1$ for ${\bf r} = {\bf r}_S$),
$\zeta_0 = R c_0$ and $\tau_0 = R^2 t_0$. The functions 
$K_{\alpha}$ and $K_{\alpha+1}$ are modified Bessel functions
\cite{AS72} with $\alpha = (d-2)/2$. From equation 
(\ref{suscept}) one obtains the asymptotic behaviour right 
at the transition point:
\begin{equation} \label{nsgauss}
N_S \sim N^{\Phi} \, \, , \qquad
\zeta_0 = \zeta_0^* \, \, , \, \, N \to \infty \, \, ,
\end{equation}
with 
\begin{equation} \label{zetastar}
\zeta_0^* = \left\{ \begin{array}{c@{\quad,\quad}r}
0 & d \le 2 \\
- (d-2) & d > 2 
\end{array} \right. \qquad .
\end{equation}
The crossover-exponent $\Phi$ 
in equation (\ref{nsgauss}) is given by \cite{EHD}
\begin{equation} \label{co}
\Phi = \frac{|d-2|}{2} \, \, , \qquad 1 \le d < 4 \, \, , 
\quad d \neq 2 \, \, ,
\end{equation}
$N_S \sim \ln N$ for $d = 2$, 
$N_S \sim N / \ln N$ for $d = 4$, and for $d > 4$
one finds $\Phi = 1$ corresponding to a first-order
transition. For $\zeta_0 < \zeta_0^*$ the finite 
fraction of adsorbed monomers scales like
\begin{equation} \label{resfrac}
\lim\limits_{N \to \infty} \frac{N_S}{N}
\sim (\zeta_0^* - \zeta_0)^{\kappa} \, \, \, , 
\quad \zeta_0 < \zeta_0^* \, \, \, ,
\end{equation}
with the exponent \cite{EHD}
\begin{equation} \label{ka}
\kappa = \frac{2 - |d-2|}{|d-2|} \, \, , \qquad 1 \le d < 4 \, \, ,
\quad d \neq 2 \, \, \, .
\end{equation}
For $d = 2$ one finds
\begin{equation} \label{fracd2}
\lim\limits_{N \to \infty} \frac{N_S}{N}
\sim e^{- 2 / |\zeta_0|} \, \, , \qquad \zeta_0 < 0 \, \, ,
\quad d = 2 \, \, ,
\end{equation}
which formally corresponds to $\kappa = \infty$.
For $d = 4$ one finds 
$\lim\limits_{N \to \infty} N_S / N
\sim - 1 / \ln (\zeta_0^* - \zeta_0)$ while for 
$d > 4$ the fraction tends to a finite value,
which reflects the fact that in this case the 
transition is of first order. Equations 
(\ref{nsgauss}) - (\ref{fracd2}) are to be compared
with equations (\ref{nscale}) - (\ref{relation})
in section \ref{subsec_general}, where the variable 
$\zeta_0^* - \zeta_0$
here corresponds to ${\cal E} - {\cal E}^*$ there. 
In particular, for given $\zeta_0 < \zeta_0^*$ the 
chain partition function $\chi(t_0)$ exhibits
a singularity for
$t_0 \searrow t_S(\zeta_0)$ where the function 
$t_S(\zeta_0)$ is determined by an analysis 
of the zero of the denominator in equation 
(\ref{suscept}) \cite{EHD}. The values 
of the exponents $\Phi$ and $\kappa$ in 
equations (\ref{co}) and (\ref{ka}) obey the scaling 
relation $\kappa = (1 - \Phi)/\Phi$
from equation (\ref{relation}). 
Note that for $d > 2$ the limit $R \to 0$ yields merely the 
trivial bulk result $\chi({\bf r}; t_0) = 1/t_0$, and hence
no phase transition for $d > 2$. Thus, in the present 
treatment it is necessary to keep the cylinder radius $R$ 
finite even though the results for $\Phi$ and $\kappa$ 
do not depend on $R$.

Finally, we note that the adsorption of a Gaussian chain 
onto a rigid rod is equivalent to the denaturation 
transition of two {\em flexible} Gaussian chains $A$ and 
$B$ if monomer $s$ of chain $A$ can only interact with 
monomer $s$ of chain $B$. This corresponds to an 
interaction of the form
$\sim \int_0^{L_0} ds \, \delta[{\bf R}_A(s) - {\bf R}_B(s)]$ 
in the partition function (\ref{pf}). 
It is easy to see that the system of two flexible Gaussian
chains with the above interaction can be mapped on 
the system of one flexible Gaussian chain 
interacting with a rigid rod, using the 
transformation \cite{CCG00,GMO01}
\begin{equation} \label{coord}
{\bf R}(s) = {\bf R}_A(s) - {\bf R}_B(s)
\, \, \, \, , \, \quad
{\bf R}_{\rm CM}(s) = \frac{1}{2} 
\left[ {\bf R}_A(s) + {\bf R}_B(s) \right]  \, \, .
\end{equation}
For Gaussian chains (and only for Gaussian chains) the degrees 
of freedom described by the centre of mass (CM) 
coordinates ${\bf R}_{\rm CM}(s)$ and the relative coordinates
${\bf R}(s)$ decouple from each other.
The relative coordinates 
${\bf R}(s)$ describe a Gaussian chain which 
interacts with the origin at ${\bf R} = 0$ in the 
$d$-dimensional subspace, while the degrees of freedom of the
remaining $D-d$ dimensions are unbounded and independent 
from the degrees of freedom of the $d$-dimensional subspace.
By virtue of this mapping, the above results 
(\ref{nsgauss}) - (\ref{fracd2}) have also been obtained
in \cite{CCG00}. 
 

\subsection{Renormalization of the field-theory 
bounded by the curved cylinder surface}
\label{subsec_ren}

We now turn to the renormalization of the $n$-vector model 
defined by equation (\ref{ham}). 
The objective is to determine the 
scaling behaviour of the renormalized chain partition function 
$\chi_{ren}(R,t,c)$ in the grand canonical ensemble in terms of 
the cylinder radius $R$ and renormalized parameters $t$ 
(conjugate to the renormalized chain length $L$) and $c$. 
The $n$-vector model in equation (\ref{ham}) can be 
dimensionally regularized and renormalized by minimal 
subtraction of poles in $\varepsilon = 4 - D$. 
The renormalizations of the bulk field 
$\vec{\Phi}({\bf r})$, ${\bf r} \in V$, and the bulk 
parameters $t_0$, $u_0$ have the same form as in the 
unbounded case, and are given by \cite{Amit,ZJ02,Diehl}
(we follow the convention of \cite{Diehl})
\begin{eqnarray} \label{rb}
\vec{\Phi} & = & Z_{\Phi}^{\,1/2} \, \vec{\Phi}_{ren} \\[2mm]
t_0        & = & \mu^2 \, Z_t \, t \, + \, t_b \\[2mm]
u_0        & = & \mu^{\varepsilon} \, 2^D \, \pi^{D/2} \, Z_u \, u
\, \, \, \, .
\end{eqnarray}
The parameters
$\vec{\Phi}_{ren}$, $t$, $u$ are renormalized counterparts
of the bare parameters $\vec{\Phi}$, $t_0$, $u_0$ in equation 
(\ref{ham}). In a regularization scheme using a large 
momentum-cutoff $\Lambda$,
the bulk renormalization factors $Z_{\Phi}$, $Z_t$, $Z_u$ 
absorb divergencies logarithmic in $\Lambda$, corresponding 
to poles in $\varepsilon$ in dimensional regularization. 
The parameter $t_b$ absorbs divergencies quadratic 
in $\Lambda$ and describes the shift of the critical 
temperature $T_c$ of the $n$-vector model due to fluctuations. 
In dimensional regularization, $t_b = 0$, and
\cite{Amit,ZJ02,Diehl}
\begin{eqnarray} \label{zb}
Z_{\Phi} & = & 1 - \frac{n+2}{36 \, \varepsilon} \, u^2 \, + \,
\Or(u^3) \\[2mm]
Z_t \, Z_{\Phi} & = & 1 + \frac{n+2}{3 \, \varepsilon} \, u \, + \,
\left[
\frac{(n+2)(n+5)}{9 \, \varepsilon^2} - \frac{n+2}{6 \, \varepsilon}
\right] u^2 \, + \, \Or(u^3) \\[2mm]
Z_u & = & 1 + \frac{n+8}{3 \, \varepsilon} \, u \, + \,
\left[
\frac{(n+8)^2}{9 \, \varepsilon^2} - \frac{3n+14}{6 \, \varepsilon}
\right] u^2 \, + \, \Or(u^3) \, \, \, .
\end{eqnarray}
The presence of the cylinder surface $S$ requires,
in addition, renormalization of the surface field 
$\vec{\Phi}|_S = \vec{\Phi}({\bf r})$, ${\bf r} \in S$,
and of the surface parameter $c_0$ in equation (\ref{ham}).
For a {\em flat} surface these additional renormalizations
are given by \cite{DD81,DD83,Diehl}
\begin{eqnarray} \label{rbsurface}
\vec{\Phi}|_S & = & (Z_{\Phi} Z_1)^{1/2} \, (\vec{\Phi}|_S)_{ren}
= Z_1^{\,1/2} \, \vec{\Phi}_{ren}|_S \\[2mm]
c_0        & = & \mu \, Z_c \, c \, + \, c_{sp} \label{rbsurface2}
\end{eqnarray}
defining the renormalized surface field $(\vec{\Phi}|_S)_{ren}$
and surface parameter $c$. The 
new renormalization factors $Z_1$ and $Z_c$ absorb 
divergencies logarithmic in $\Lambda$ which occur
at a flat surface, corresponding 
to poles in $\varepsilon$ in dimensional regularization.
The parameter $c_{sp}$
absorbs divergencies linear in $\Lambda$ and describes 
the shift of the multicritical point due to 
fluctuations (compare figure \ref{fig_pd},
$- c \sim {\cal E}$).
In dimensional regularization, $c_{sp} = 0$,
and $Z_1$ and $Z_c$ are given by 
\cite{DD81,DD83,Diehl}
\begin{eqnarray} \label{zs1}
Z_1 & = & 1 + \frac{n+2}{3 \, \varepsilon} \, u \, + \,
\left[
\frac{(n+2)(n+5)}{9 \, \varepsilon^2} - \frac{n+2}{3 \, \varepsilon}
\right] u^2 \, + \, \Or(u^3) \\[2mm]
Z_c & = & 1 + \frac{n+2}{3 \, \varepsilon} \, u \, + \,
\left[
\frac{(n+2)(n+5)}{9 \, \varepsilon^2} + \frac{n+2}{36 \, \varepsilon}
\, (1 - 4 \pi^2) \right] u^2 \, + \, \Or(u^3) \, \, \, . \label{zs2}
\end{eqnarray}
Equations (\ref{rbsurface}) and (\ref{rbsurface2}) 
hold for a flat 
surface. As shown by McAvity and Osborn \cite{AO93},
the required renormalizations become modified if 
the surface $S$ is curved, like in the present case.
While the renormalization of the surface field
$(\vec{\Phi}|_S)_{ren}$ 
remains unchanged, the renormalization of the surface 
parameter $c_0$ requires an additional, additive term 
that depends on the mean curvature \cite{AO93}:
\begin{equation} \label{rs1}
c_0 = \mu \, Z_c \, c \, + \, \frac{d-1}{R} \, 
{\cal C}(u,\varepsilon)
\end{equation}
or, with $\zeta_0 = R c_0$ and $\zeta = \mu R c$,
\begin{equation} \label{rs1short}
\zeta_0 = Z_c \, \zeta \, + \, (d-1) \, 
{\cal C}(u,\varepsilon) \, \, .
\end{equation}
$Z_c$ is the same renormalization factor
as in equation (\ref{rbsurface2}) for a flat surface
and ${\cal C}(u,\varepsilon)$ to second order in $u$ 
can be deduced from reference \cite{AO93}:
\begin{equation} \label{c}
{\cal C}(u,\varepsilon) = \frac{n+2}{9}
\left\{
\frac{u}{\varepsilon} +
\left[\frac{n+5}{3 \, \varepsilon^2} + 
\frac{n+1 - 4 \pi^2}{12 \, \varepsilon} \right] u^2
\right\}  \, + \, \Or(u^3) \, \, \, .
\end{equation}

To proceed, we consider the two-point correlation function
\begin{equation} \label{intcfrage}
\langle \Phi_1({\bf r}_S) \, \Phi_1({\bf r}') \rangle 
= G({\bf r}_S, {\bf r}', R; t_0, \zeta_0, u_0)
\, \, ,
\end{equation}
where ${\bf r}_S$ is located on the cylinder 
surface $S$ and 
${\bf r}' \in V$. From $G$, the chain partition 
function $\chi(R,t_0,\zeta_0,u_0)$ follows by an 
integration over ${\bf r}'$, see equation (\ref{intc});
by symmetry, $\chi$ does not depend on ${\bf r}_S$.
The renormalization group (RG) equation for the 
renormalized counterpart $G_{ren}$ of $G$ follows 
in the standard way, using the relation
\begin{equation} \label{gg}
G_{ren}({\bf r}_S, {\bf r}', R; t, \zeta, u; \mu) =
Z_{\Phi}^{\,-1} Z_1^{\,-1/2} \, 
G({\bf r}_S, {\bf r}', R; t_0, \zeta_0, u_0)
\end{equation}
and the fact that $G$ does not depend on $\mu$: 
$\mu \partial_{\mu} G|_{\mu=0} = 0$. This 
leads to the RG equation
\begin{equation} \label{rg}
\left[
{\cal D}_{\mu} + \eta_{\Phi} + \frac{1}{2} \, \eta_1 + 
\mu \frac{\partial \zeta}{\partial \mu}\,\Big|_{\mu=0} \, \partial_{\zeta}
\right] G_{ren} \, = \, 0 
\end{equation}
where we have used the abbreviation
\begin{equation} \label{d}
{\cal D}_{\mu} = \mu \partial_{\mu}|_{\mu=0} +
\beta(u) \partial_u - (2 + \eta_t) \, t \, \partial_t
\end{equation}
with
\begin{equation} \label{beta}
\beta(u) = \mu \partial_{\mu} u|_{\mu=0}
\end{equation}
and the exponent functions
$\eta_i(u) = \mu \partial_{\mu} \ln Z_i|_{\mu=0}$
with $i = \Phi, 1, t$.
The new feature of the RG equation (\ref{rg}) 
generated by the surface curvature is the function  
\begin{equation} \label{zeta}
\mu \frac{\partial \zeta}{\partial \mu}\,\Big|_{\mu=0} =
- \, \eta_c \left[
\zeta + \frac{d-1}{3} + \frac{n (d-1)}{18} \, u \, + \, \Or(u^2) 
\right]
\end{equation}
where
\begin{eqnarray} \label{etac}
\eta_c(u) & = & \mu \partial_{\mu} \ln Z_{c\,}|_{\mu=0} \\[1mm]
       & = & - \, \frac{n+2}{3} \, u \, + \, 
\frac{n+2}{18} \, (4 \pi^2 - 1) u^2  \, + \, \Or(u^3) \nonumber
\end{eqnarray}
is the exponent function for $c$ corresponding to a {\em flat} 
surface, with $Z_c$ from equation (\ref{zs2}) \cite{Diehl}.
A necessary condition for the two-point correlation function $G$ 
to be scale invariant 
(SI) is that the right hand side of equation (\ref{zeta}) vanishes, 
which is either the case for $u = 0$ 
(Gaussian model) or if $\zeta$ takes the value for which 
the square bracket in equation (\ref{zeta}) vanishes:
\begin{equation} \label{van}
\zeta = \zeta_{\rm SI} \equiv - \, \frac{d-1}{3} \, + \, \Or(u) \, \, .
\end{equation}
Note that the value of $\zeta_{\rm SI}$ to leading order in $u$,
$\zeta_{\rm SI} = - (d-1)/3$, is {\em different} from the value 
$\zeta_0^*$ in equation (\ref{zetastar}) corresponding to
the onset of the adsorption transition in the Gaussian
model. It is also interesting to note that for a sphere,
corresponding to $d = D = 4 - \varepsilon$, 
the value of $\zeta_{\rm SI}$ 
to leading order in $u$, $\zeta_{\rm SI} = - 1$, coincides with
the value $\zeta_{\rm CI} = - (d-2)/2$ for which the Gaussian 
two-point correlation function at $t_0 = 0$
is conformal invariant (CI) 
\cite{ER95,rem}. However, within the $\Phi^4$-model the special
value $\zeta_{\rm SI}$ is already fixed by scale invariance, 
whereas in the Gaussian model the value $\zeta_{\rm CI}$ is 
only fixed if one requires conformal invariance.

By solving the RG equation (\ref{rg}) for the 
two-point function 
in the standard way using characteristics and integrating 
over ${\bf r}'$, one derives the general scaling behaviour 
of the chain partition function in the grand canonical 
ensemble:
\begin{equation} \label{scalbeh}
\chi_{ren}(R, t, \zeta, u; \mu)
\sim t^{-\gamma_1} \,
\Theta\left(
\alpha \mu R t^{\nu}, \beta \Delta \zeta t^{\nu - \Phi_{\rm flat}} \right)
\end{equation}
where 
\begin{equation} \label{variable}
\Delta \zeta = \zeta - \zeta_{\rm SI}
\end{equation}
with $\zeta_{\rm SI}$ from equation (\ref{van}).
The exponent $\nu = (2 + \eta_t^*)^{-1}$ is a bulk critical 
exponent, while $\Phi_{\rm flat} = \nu (1 + \eta_c^*)$ 
(the crossover exponent) and 
$\gamma_1 = \nu (2 - \eta_{\Phi}^* - \eta_1^* / 2)$
are critical exponents associated with a {\em flat} surface 
\cite{Diehl}.
The exponent functions $\eta_i(u)$ for $i = t, c, \Phi, 1$
are defined below equation (\ref{beta}) and in 
equation (\ref{etac}).
The values $\eta_i^*$ are the values 
of the exponent functions at the fixed point $u^*$.  
The constants $\alpha$ and $\beta$
in equation (\ref{scalbeh}) are nonuniversal prefactors,
while the function $\Theta(x,y)$ is a universal scaling 
function.

Equation (\ref{scalbeh}) is in line with the 
Gaussian model (where $\nu = \Phi_{\rm flat} = 1/2$
and $\gamma_1 = 1$):
\begin{equation} \label{scalbehgauss}
\chi(R, t_0, \zeta_0)
\sim t_0^{-1} \,
\Theta_0(R t_0^{1/2}, \zeta_0) \, \, \, , \qquad \mbox{Gaussian model} \, \, ,
\end{equation}
compare equation (\ref{suscept}).
Finally, equation (\ref{scalbeh}) can be compared with 
the corresponding behaviour for a flat surface:
\begin{equation} \label{scalbehplanar}
\chi_{ren}(t, c, u; \mu)
\sim t^{-\gamma_1} \,
\Theta_{\rm flat}(\beta' c \, t^{- \Phi_{\rm flat}})
\, \, \, , \qquad \mbox{flat surface} \, \, .
\end{equation}
Note that this limit can be obtained from equation 
(\ref{scalbeh}) by rewriting the second scaling 
argument as 
$\Delta \zeta t^{\nu - \Phi_{\rm flat}} = 
R \, t^{\nu} \cdot \Delta c \, t^{- \Phi_{\rm flat}}$.
Equation (\ref{scalbehplanar}) then follows as the 
limit $R \to \infty$ of the scaling form
$\chi_{ren} \sim t^{-\gamma_1} \,
\widetilde{\Theta}(\alpha' \mu R \, t^{\nu}, 
\beta' \Delta c \, t^{- \Phi_{\rm flat}})$,
where $\Delta c = c + \Or(1/R)$.

Let us come back to equation (\ref{scalbeh}).
According to recent estimates one has $\Phi_{\rm flat} = 0.518$ 
\cite{DS94} and $\nu = 0.588$ \cite{ZJ02} for $n=0$
in $D=3$ so that the exponent $\nu - \Phi_{\rm flat}$ in 
equation (\ref{scalbeh}) is small but positive. 
From a naive point of view this would imply that 
the scaling variable $\Delta \zeta$ in the second 
scaling argument of $\Theta$ is {\em irrelevant}
and could be omitted from the outset;
however, one should keep in mind that the radius $R$ 
in the first scaling argument is also irrelevant in
principle. Now, the relevant question is 
whether the scaling function $\Theta(x,y)$ exhibits 
a singularity on a certain subset of $(x,y)$,
corresponding to the polymer adsorption transition; 
compare the related discussion below equation 
(\ref{scalex}) and below equation (\ref{fracd2}).
In fact, we expect this singularity to occur for 
$y = y_S(x) < 0$, where now $y_S(x)$ is a function 
of the first scaling variable $x = \alpha \mu R t^{\nu}$. 
Note that the scaling function $\Theta_0(x,y)$ in equation 
(\ref{scalbehgauss}), corresponding to the Gaussian 
model, exhibits this kind of singularity indeed;
compare section \ref{subsec_gauss}.
Thus, in the present description, the adsorption 
transition is characterized by a
balance of the two irrelevant variables
$\Delta \zeta$ and $R$. In this sense
the scaling variable $\Delta \zeta$ 
can be considered as a dangerously 
irrelevant variable.

Finally we note that
equation (\ref{scalbeh}) implies a corresponding 
scaling form of the chain partition function with
fixed chain length $L$:
\begin{equation} \label{scalbehz}
Z_{ren}(R, L, \zeta, u; \mu)
\sim L^{\gamma_1 - 1} \,
\Psi\left(
\widetilde{\alpha} \mu R L^{-\nu}, 
\widetilde{\beta} \Delta \zeta L^{-(\nu - \Phi_{\rm flat})} \right) \, \, ,
\end{equation}
with (different) nonuniversal prefactors 
$\widetilde{\alpha}$ and $\widetilde{\beta}$
and a universal scaling function $\Psi$.
From equations (\ref{scalbeh}) and (\ref{scalbehz}) the 
number of adsorbed monomers $N_S$
for $\Delta \zeta = 0$ and the finite fraction of adsorbed
monomers $N_S / N$ for $\Delta \zeta < 0$ can
be derived as outlined in section \ref{subsec_general}
in principle. However,
the thin rod limit $R \to 0$ corresponds to singular 
limits of the scaling functions $\Theta$ and $\Psi$ which 
are rather difficult to obtain. At least it is easy to see
that the exponents $\Phi$ and $\kappa$ defined in equations
(\ref{nscale}) and (\ref{resfracintro}) are {\em universal}, 
using the fact that $\Theta$ and $\Psi$ are universal 
scaling functions. To proceed, in the next section we 
use a different 
method to obtain estimates for the exponents 
$\Phi$ and $\kappa$ for a cylinder in $D = 3$.


\subsection{Estimates for the exponents $\Phi$ and $\kappa$
by using the additivity of co-dimensions}

In this section we obtain estimates for the exponents 
$\Phi$ and $\kappa$ for a cylinder in $D=3$ introduced 
in equations (\ref{nscale}) and (\ref{resfracintro}) by 
means of an interpolation procedure between two
known cases.  
Firstly, in the Gaussian model one has $\Phi = 1/2$ 
for $d = 3$, see equation (\ref{co}), corresponding to 
the point $(d,D) = (3,4)$ in figure \ref{fig_cyl}. 
Likewise, $\kappa = 1$ from the scaling relation 
$\kappa = (1 - \Phi)/\Phi$, or from equation (\ref{ka}),
for $(d,D) = (3,4)$.
Secondly, one has $\Phi = 0$ on the whole line 
$v^{-1}(D) - d = 0$ in figure \ref{fig_cyl}.
This result can be obtained by using the
{\em co-dimension additivity theorem}, stating
that the co-dimension of the intersection points
of two objects of dimensions 
$D_1$ and $D_2$ is given by the sum of their 
co-dimensions:
$D - D_{int} = (D - D_1) + (D - D_2)$, i.e.,
\begin{equation} \label{codim}
D_{int} = D_1 + D_2 - D \, \, \, .
\end{equation}
For example, two-dimensional surfaces generically intersect 
along curves in $D = 3$ ($D_{int} = 2 + 2 - 3 = 1$) 
and only at isolated points in $D = 4$  
($D_{int} = 0$).
Equation (\ref{codim}) is also expected to hold if 
one or both objects are fractal. 
In the present case, one object is a self-avoiding 
random walk with 
fractal (Hausdorff) dimension $v^{-1}$ and the other 
one is a ``generalized cylinder'' with co-dimension $d$
(see figure \ref{fig_cyl}); the dimension of 
intersection points of these two objects is thus 
given by 
\begin{equation} \label{codimen}
D_{int} = \nu^{-1}(D) - d \, \, \, .
\end{equation}
In figure \ref{fig_cyl}, the line $D_{int} = 0$ as a function of 
$d$ and $D$ is shown as the blue dashed line. An unbounded, 
free, self-avoiding random walk does not intersect with 
``generalized cylinders'' located above the dashed line,
apart from exceptional cases. In this sense, 
``generalized cylinders''
above the dashed line are irrelevant perturbations 
for a free, self-avoiding random walk. 
Now, ``generalized cylinders'' located right on the 
line $D_{int} = 0$ correspond to marginal cases:  
An unbounded, free, self-avoiding random walk {\em does}
intersect with ``generalized cylinders'' located on the 
dashed line, but only at isolated points. We thus expect 
that the number of intersecting monomers $N_S$ grows with 
$N$ for $N \to \infty$, but only logarithmically, i.e.,
$N_S \sim \ln N$, which implies $\Phi = 0$; 
compare the case $d = 2$ for the Gaussian model discussed in 
section \ref{subsec_gauss}, and compare the case 
${\cal E} = {\cal E}^*$ with $\Phi = 0$ in section
\ref{subsec_general}.
It should be noted that this argument only applies to 
{\em unperturbed} random walks, and does not make 
any statement for walks that interact with the body.

Thus, the values of the exponent
$\Phi$ at the end points of the green
line in figure \ref{fig_cyl} are available. This can be
used to obtain an estimate for $\Phi$ for an ordinary 
cylinder in $D = 3$ as follows.
The shape of the dashed line in figure 
\ref{fig_cyl} is known quite accurately by means of the 
$\varepsilon$-expansion of $\nu(D)$ in conjunction 
with the exact value $\nu = 3/4$ for $n = 0$ in 
$D = 2$ \cite{Amit,ZJ02}. Thus, one may
estimate $\Phi$ for a cylinder in $D=3$,
located at the point 
$(d,D) = (2,3)$ in figure \ref{fig_cyl},
by means of a
linear interpolation between the known values 
of $\Phi$ at the end points of the green line
(compare references \cite{HD99,H2000}). 
In this way we find 
for a cylinder in $D = 3$ the estimates,
using equation (\ref{relation}),
\begin{equation} \label{exp}
\Phi \simeq \frac{1}{6} \, \, \, , \qquad
\kappa = \frac{1 - \Phi}{\Phi} \simeq 5 \, \, \, .
\end{equation}
Since these exponents are universal and do not depend on
the cylinder radius $R$, they are also expected to hold
for a rigid rod with vanishing radius, or for a line
of lattice sites in a numerical simulation of this system.


\section{Conclusion}

We have investigated the adsorption transition 
of a long flexible self-avoiding polymer chain 
onto a rigid thin rod by field-theoretical 
methods.
The rod is endowed with a short-ranged adsorption 
energy ${\cal E}$ for the chain monomers so that,
on increasing ${\cal E}$, at some threshold 
value ${\cal E}^*$ the chain undergoes a transition 
from an unbound state to a bound state, as shown in
figure \ref{fig_rod}. The main results and 
remaining questions are summarized below. 

\begin{itemize}

\item[1)] By means of general scaling arguments we
obtained the scaling relation (\ref{relation}) for 
the exponents $\Phi$ and $\kappa$ defined 
in equations (\ref{nscale}) and (\ref{resfracintro}), 
and the phase diagrams shown in figure \ref{fig_pd}
in terms of the number of chain monomers $N$, the
chemical potential $\mu$ conjugate to $N$, and the 
adsorption energy ${\cal E}$.

\item[2)] By representing the rod by a cylinder of 
finite radius $R$ we could use available results for 
field theories with curved boundaries \cite{AO93}; 
see figure \ref{fig_curved}. By using renormalization
group arguments, we derived
the scaling behaviour of the chain 
partition function in the grand canonical ensemble,
equation (\ref{scalbeh}), and in the canonical ensemble,
equation (\ref{scalbehz}), where $L \sim N$ and
$t \sim \mu$. Notable features of the scaling 
results are the distinct form of the scaling variable 
$\zeta \propto R c$, where the parameter $c$ is 
related to the surface potential
for chain monomers, and the curvature-induced shift of
$\zeta$ in equation (\ref{variable}) with $\zeta_{\rm SI}$
from equation (\ref{van}). It also follows that the
exponents $\Phi$ and $\kappa$ introduced in equations 
(\ref{nscale}) and (\ref{resfracintro}) are 
universal.

\item[3)] Because the cylinder radius $R$ enters 
the scaling functions $\Theta$ and $\Psi$ in equations 
(\ref{scalbeh}) and (\ref{scalbehz}) explicitly
it is difficult to obtain the universal exponents $\Phi$ 
and $\kappa$ directly from them.
Therefore we used the co-dimension additivity theorem 
in conjunction with an interpolation procedure,
as shown in figure \ref{fig_cyl}, to obtain the 
estimates for $\Phi$ and $\kappa$ in equation 
(\ref{exp}). The check of these exponents and the
scaling relation (\ref{relation}) is a possible 
starting point for numerical simulations of this
system.

\item[4)] It would be interesting to introduce new
methods to derive the exponents $\Phi$ and $\kappa$, 
possibly avoiding the introduction of a finite cylinder 
radius $R$ from the outset.

\item[5)] It would also be interesting to explain 
the relation between $\zeta_0^*$ and $\zeta_{\rm SI}$ 
discussed below equation (\ref{van}).

\end{itemize}


\ack I would like to thank C~v~Ferber for useful 
correspondence.


\appendix
\section{Mapping of the polymer system on the $n$-vector
model}
\setcounter{section}{1}
\label{sec_app}

The $\rho^2({\bf r})$ interaction in equation (\ref{pf}) 
can be linearized by means of a Gaussian transformation
\cite{C75,CJ90}. 
This procedure makes use of the Gaussian integral 
\begin{equation} \label{gauss}
\int {\cal D}X \exp\left[- \frac{1}{2} \, X^T A X + b^T X \right] =
\left( \det \frac{A}{2 \pi} \right)^{-1/2} \,
\exp\left[\frac{1}{2} \, b^T A^{-1} \, b \right]
\end{equation}
where $X$ is a vector with discrete or continuous indices and 
the symmetric matrix $A$ must have a positive definite real 
part. Using $X({\bf r}) \propto i \sigma({\bf r})$ with purely 
imaginary $\sigma({\bf r})$, the matrix
$\displaystyle{A({\bf r},{\bf r}') = 
\frac{3}{u} \, \delta({\bf r}-{\bf r}')}$,
and $b({\bf r}) = i \rho({\bf r})$, one finds
\begin{eqnarray} \label{aux}
& & \exp \left\{ - \frac{u}{6} 
\int d^Dr \, \rho^2({\bf r}) \right\} \\[2mm]
& & = \int {\cal D} \sigma \,
\exp\left[ \frac{3}{2 u} \int d^D r \, \sigma^2({\bf r}) \, -
\int d^Dr \, \sigma({\bf r}) \rho({\bf r}) \right] \, \, . \nonumber
\end{eqnarray}
Note that $A$ is positive definite due to our 
assumption  $u > 0$.
Inserting (\ref{aux}) in (\ref{pf}) yields
\begin{eqnarray} \label{pf2}
& & Z^{(2)}({\bf r}, {\bf r}'; L) =  
\int {\cal D} \sigma \,
\exp\left[ \frac{3}{2 u} 
\int d^D r \, \sigma^2({\bf r}) \right] \\[2mm]
& & \, \, \times \int_{\bf r}^{{\bf r}'} {\cal D} {\bf R}
\, \exp \left\{ - \frac{1}{4}
\int_{0}^{L} ds \left(\frac{d {\bf R}}{ds}\right)^2 
- \int d^Dr \, 
\left[ V({\bf r}) + \sigma({\bf r}) \right]
\rho({\bf r}) \right\} \nonumber
\end{eqnarray}
The $\rho^2({\bf r})$ interaction in equation 
(\ref{pf}) has been replaced by the interaction of $\rho({\bf r})$
with an external, fluctuating potential $\sigma({\bf r})$.
The second line of equation (\ref{pf2}) can be 
interpreted as the path integral representation of 
the evolution operator 
$\langle {\bf r}' \, | \, e^{- L \hat{H} } | \, {\bf r} \, \rangle$
in imaginary time $s$ of a quantum-mechanical particle
with Hamiltonian 
$\hat{H} = - \Delta + V({\bf r}) + \sigma({\bf r})$.
The Laplace transform of this evolution operator with
respect to $L$ yields the resolvent
\begin{equation} \label{lapop}
\int_0^{\infty} dL \, e^{- t L} \, 
\langle {\bf r}' \, \bigg| \, e^{- L \hat{H} } \bigg| \, {\bf r} \, \rangle
= \left\langle {\bf r}' \, 
\Big| \frac{1}{- \Delta + t + V({\bf r}) + \sigma({\bf r})} \Big| 
\, {\bf r} \right\rangle \, \, .
\end{equation}
The resolvent can be represented in the standard way
by the two-point correlation
function of an $n$-component field 
$\vec{\Phi} = (\Phi_1, \ldots, \Phi_n)$ in the limit 
$n \to 0$. The result is
\begin{eqnarray} \label{ev}
Z^{(2)}({\bf r}, {\bf r}'; L) & = & 
\int {\cal D} \sigma
\exp\left[ \frac{3}{2 u} 
\int d^D r \, \sigma^2({\bf r}) \right] \\[2mm]
& \times & {\cal L}_{t \to L} \, \lim\limits_{n \to 0} \,
\int {\cal D} \vec{\Phi} \,
\Phi_1({\bf r}) \, \Phi_1({\bf r}') \, e^{- S\{\vec{\Phi}\}} \nonumber
\end{eqnarray}
with the action
\begin{equation} \label{actionev}
S\{\vec{\Phi}\} = \int d^Dr \left[ 
\frac{1}{2} (\nabla \vec{\Phi})^2 + 
\frac{t}{2} \, \vec{\Phi}^2 + 
\frac{1}{2} \, \left[ \sigma({\bf r}) + V({\bf r}) \right]
\vec{\Phi}^2 \right] \, \, \, .
\end{equation}
The Gaussian integration in equation (\ref{ev}) can be carried
out using equation (\ref{gauss}) with the same
$X({\bf r}) \propto i \sigma({\bf r})$ 
and matrix $\displaystyle{A({\bf r},{\bf r}') = 
\frac{3}{u} \, \delta({\bf r}-{\bf r}')}$ as before, and now
$\displaystyle{b({\bf r}) = \frac{i}{2} \, \vec{\Phi}^2({\bf r})}$.
This leads to equations (\ref{res}) - (\ref{action}).


\end{document}